\documentclass[amsmath,amssymb,draftaps,showpacs,twocolumn,superscriptaddress]
{revtex4-1}
\usepackage{graphicx}
\usepackage{dcolumn}
\usepackage{bm}
\usepackage{braket}
\usepackage{color}
\definecolor{mycolor1}{rgb}{0,0,1}

\usepackage{ulem}

\begin{document}

\preprint{APS/123-QED}

\title{Different metamagnetism between paramagnetic Ce and Yb isomorphs}

\author{Atsushi Miyake}
\thanks{miyake@issp.u-tokyo.ac.jp}
\affiliation{Institute for Solid State Physics, The University of Tokyo, Kashiwa, Chiba 277-8581, Japan}

\author{Yoshiaki Sato}
\affiliation{Institute for Solid State Physics, The University of Tokyo, Kashiwa, Chiba 277-8581, Japan}

\author{Masashi Tokunaga}
\affiliation{Institute for Solid State Physics, The University of Tokyo, Kashiwa, Chiba 277-8581, Japan}

\author{Jumaeda Jatmika}
\affiliation{Department of Physics, Shizuoka University, Suruga, Shizuoka, 422-8529, Japan}

\author{Takao Ebihara}
\affiliation{Department of Physics, Shizuoka University, Suruga, Shizuoka, 422-8529, Japan}

\date{\today}

\begin{abstract}

To solve the puzzle of metamagnetic phenomena in heavy fermion systems, we have compared paramagnetic isostructural Ce and Yb systems, CeNi$_2$Ge$_2$ and YbNi$_2$Ge$_2$, both of which are located near a magnetic instability.
The most intriguing result is discovery of a metamagnetic-like transition in both systems from magnetization measurements in a pulsed magnetic fields.
This is the first observation of the metamagnetism for isomorphic Ce and Yb paramagnetic systems.
Similar to other metamagnets, the metamagnetic transition fields for both compounds are well scaled by the temperature $T_{\chi}^{\rm max}$, at which the magnetic susceptibility shows a maximum.
In addition, for CeNi$_2$Ge$_2$ a peak of nonlinear susceptibility $\chi_3$ appears at approximately $T_{\chi}^{\rm max}/2$, as for other heavy-fermion metamagnets.
In contrast, YbNi$_2$Ge$_2$ shows only a sign change for $\chi_3$ at $T_{\chi}^{\rm max}$, as observed in itinerant metamagnets located near the ferromagnetic critical point.
The metamagnetism of CeNi$_2$Ge$_2$ corresponds to a typical Kondo lattice system, whereas that of YbNi$_2$Ge$_2$ is similar to the nearly ferromagnetic itinerant systems. 
Other possibilities for the metamagnetic behavior of YbNi$_2$Ge$_2$ are also discussed. 
\end{abstract}

\maketitle

\section{Introduction}
Metamagnetic phenomena in strongly correlated itinerant electron systems have attracted much attention for a long time.
Since the discovery of a nonlinear increase in magnetization $M$ of CeRu$_2$Si$_2$ at characteristic fields of $H_m\sim$7.8~T along the tetragonal $c$-axis \cite{Haen1987}, extensive experimental and theoretical studies have been performed.  
Metamagnetism of CeRu$_2$Si$_2$ is regarded as a crossover rather than as a phase transition, and hence referred to as a pseudo-metamagnetic transition.
When the system is tuned to the antiferromagnetic (AFM) phase by expanding its volume by chemical pressure, the first-order metamagnetic transition from AFM to the polarized paramagnetic (PPM) phase takes place at $H_c$ \cite{Fisher1991}. 
In crossing the critical pressure ($p_{\rm c}$) at which the AFM transition temperature $T_{\rm N}$ suppressed to zero, the metamagnetic transition for CeRu$_2$Si$_2$ becomes crossover, e.g. $H_c\sim H_m$ \cite{Fisher1991, Flouquet2005}.
Another example of metamagnetism is field-induced paramagnetic (PM) to ferromagnetic (FM) transition for the itinerant electron systems located near the FM critical point \cite{Yamada1993, Goto2001}. 
When the system is in the PM phase beyond the FM critical endpoint, the metamagnetic transition changes to crossover, as observed in UCoAl \cite{Goto2001, Aoki2011}.
In many itinerant electron systems, such as PM heavy-fermion and nearly FM systems, metamagnetism only appears below $T_{\chi}^{\rm max}$, where the magnetic susceptibility $\chi$ is at a maximum.
In addition, $H_m$ is known to be proportional to $T_{\chi}^{\rm max}$ \cite{Aoki2013, Goto2001}.
These facts indicate that metamagnetism and the maximum in $\chi$ are dominated by a single energy scale, i.e., they have the same origin.

To reveal the more details of metamagnetism in heavy-fermion systems, we focused on certain aspects of the well-studied ThCr$_2$Si$_2$-type tetragonal Ce and Yb systems as the following aspects.
In the tetragonal symmetry, the crystalline electric field (CEF) split the $J =$~5/2~(7/2) of Ce$^{3+}$ (Yb$^{3+}$) into three (four) Kramers doublets.
When the CEF splitting energy $\Delta_{\rm CEF}$, typically of the order of 100-200~K, is larger than the Kondo temperature $T_{\rm K}$, the degeneracy of the doublet ground state must be resolved by forming the magnetic order or heavy-fermion state.
The balance between $T_{\rm K}$ and $\Delta_{\rm CEF}$ can be tuned by composition or the external parameters, such as pressure and doping. 
From the literature \cite{Aoki2013}, only the hexagonal and cubic compounds of the PM Yb-systems were known to show metamagnetism, for example, YbCuAl \cite{Hewson1983}, YbAgCu$_4$ \cite{Graf1995}, YbCu$_5$ \cite{Tsujii1997}, and  Yb$T_2$Zn$_{20}$ ($T$~=~Co, Rh and Ir) \cite{Yoshiuchi2009, Hirose2011}.
In this context, the discovery of new examples exhibiting metamagnetism among the PM Yb systems having tetragonal symmetry would be desirable.
Given the above issues, we focus on the Ce and Yb isomorphs, CeNi$_2$Ge$_2$ and YbNi$_2$Ge$_2$.
Both compounds crystalize in the tetragonal ThCr$_2$Si$_2$-type structure and have PM ground states like CeRu$_2$Si$_2$.
It is quite rare that both Ce and Yb isomorphs have a PM ground state and therefore this comparison may shed light on the difference or similarity between the 4$f$-electron and hole analogues. 
 
CeNi$_2$Ge$_2$ is believed to be located near the AFM instability.
At low temperature, the electrical resistivity, specific heat and magnetic susceptibility deviate from the Fermi-liquid description \cite{Aoki1997,Gegenwart1999}.
For example, the low-temperature specific heat divided by temperature exhibits a $-\sqrt{T}$ dependence with large value of the coefficient of electronic specific heat $\gamma$~=~350-450~mJ/mol~K$^2$~~\cite{Gegenwart1999, Knopp1988,  Aoki1997}.
The thermal Gr$\ddot{\rm u}$neisen parameter diverges as $T$~$\rightarrow$~0 \cite{Kuchler2003}.
Large $\gamma$-value and  the diverging of Gr$\ddot{\rm u}$neisen parameter recall to mind CeRu$_2$Si$_2$ \cite{Lacerda1989}.
Both CeNi$_2$Ge$_2$ and CeRu$_2$Si$_2$ have the large values exceeding 100 as encountered in many heavy-fermion systems \cite{deVisser1989, Kuchler2007}.
The temperature dependence of $\chi$ for $H~||~c$-axis features a broad maximum at $T_{\chi}^{\rm max}\sim$~28~K \cite{Fukuhara1996} for CeNi$_2$Ge$_2$, whereas $T_{\chi}^{\rm max}\sim$~10~K for CeRu$_2$Si$_2$ \cite{Haen1987}.
The Pd substitutions at the Ni sites of CeNi$_2$Ge$_2$ induce AFM ordering, which indicates proximity to an AFM phase. \cite{Fukuhara1998, Knebel1999, Fukuhara2002}.
CeNi$_2$Ge$_2$ was reported to exhibit metamagnetism at $H_m\sim$~42~T in free powdered samples \cite{Fukuhara1996} and $\sim$43~T in oriented powdered samples \cite{Gegenwart2003}.
Because of the relatively high $H_m$, the details of metamgenetism for CeNi$_2$Ge$_2$ are still unclear.

YbNi$_2$Ge$_2$ has the relatively large $\gamma$-value of 136 mJ/mol K$^2$ \cite{Budko1999} and has an intermediate Yb valence of $\sim$~2.8 at low temperature \cite{Yamaoka2010}.  
Interestingly, $\chi$ shows a broad maximum at approximately 50~K, for both $H~||~a$ and $c$ \cite{Budko1999}.
Magnetic ordering was observed above $p_c$=~5~GPa \cite{Knebel2001}, at which the Yb valence remains non-integer \cite{Yamaoka2010}.
Although the magnetic structure above $p_c$ is still not known, a FM interaction is indicated from the magnetoresistance \cite{Knebel2001}.
This fact infers that YbNi$_2$Ge$_2$ is located near a FM critical point.
This is strikingly different from CeNi$_2$Ge$_2$, which is located near an AFM instability.
Moreover, their magnetic easy directions are different, as seen in the susceptibility curves (see Fig.~\ref{MT}).
From scaling, $H_{m}$~$\sim$~$T_{\chi}^{\rm max}$, the metamagnetic behavior for both CeNi$_2$Ge$_2$ and YbNi$_2$Ge$_2$ is expected to be captured using a pulsed magnetic field.

In this paper, we compare the metamagnetic behavior of paramagnetic CeNi$_2$Ge$_2$ and YbNi$_2$Ge$_2$ obtained from magnetization measurements for fields up to 56 T using a pulsed magnet.
Both are located near their respective magnetic critical point: AFM for CeNi$_2$Ge$_2$ and FM for YbNi$_2$Ge$_2$;
the magnetic anisotropy is also different, i.e., the easy magnetization $c$-axis for the former and the easy basal plane for the latter.
The main observation here is that both compounds feature a pseudo-metamagnetic magnetization anomalies when the field is applied along the easy magnetization axis or plane.  
In particular, YbNi$_2$Ge$_2$ might be the first example of a PM Yb compound with tetragonal symmetry exhibiting a metamagnetic behavior. 
Differences appear in their temperature evolutions of magnetization and nonlinear magnetic susceptibility.

\section{Experiment}

Single crystals of CeNi$_2$Ge$_2$ were prepared by the Czochralski method, and those of YbNi$_2$Ge$_2$ were grown by the In-flux method \cite{Budko1999}.
The temperature dependence of magnetization at 0.1~T is measured using a commercial SQUID magnetometer.
Pulsed-magnetic fields up to 56~T were applied using a non-destructive magnet with typical durations of $\sim$ 36~ms installed at the International MegaGauss Science Laboratory of Institute for Solid State Physics at the University of Tokyo.
Magnetization in a pulsed field was measured by conventional induction method, using coaxial pick-up coils.

\section{Results and discussions}

\begin{figure}[t]
   \begin{center}
      \includegraphics[width=90mm]{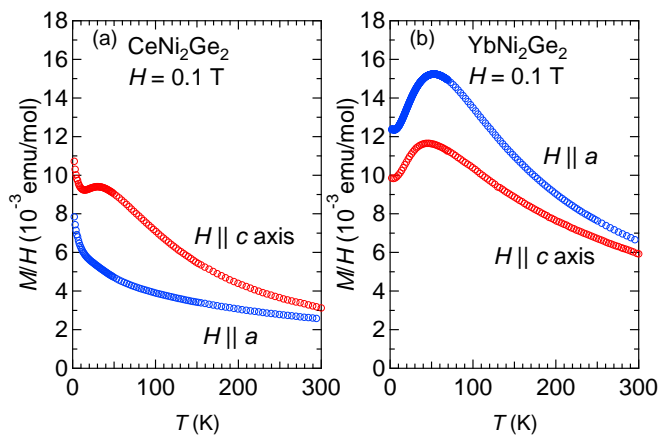}
   \end{center}
   \caption{(color online) Temperature dependence of the magnetic susceptibility $M/H$ of (a) CeNi$_2$Ge$_2$ and (b) YbNi$_2$Ge$_2$ in magnetic fields of 0.1~T applied along $a$ and $c$ axes. 
   }
   \label{MT}
\end{figure}

\begin{figure}[t]
   \begin{center}
      \includegraphics[width=90mm]{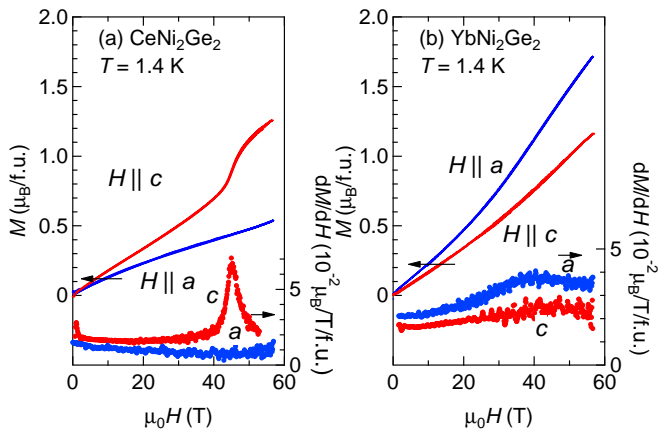}
   \end{center}
   \caption{(color online) Magnetic field dependence of magnetization at 1.4~K of (a) CeNi$_2$Ge$_2$ and (b) YbNi$_2$Ge$_2$ for $H||a$~and $c$ axes.
   The differential susceptibility d$M$/d$H$ for each is are also presented.
}
   \label{MH}
\end{figure}

Figure \ref{MT} presents the temperature dependence of the magnetic susceptibility $M/H$ for applied magnetic fields of 0.1~T along the tetragonal $a$- and $c$-axes for CeNi$_2$Ge$_2$ and YbNi$_2$Ge$_2$.   
The overall trends are in good agreement with the previous reports \cite{Fukuhara1996, Budko1999, Yamaoka2010}.
From the Curie-Weiss fit above 100~K, the effective moment of CeNi$_2$Ge$_2$, estimated to be $\sim$3.0~$\mu_{\rm B}$ for the both directions, is slightly larger than a previous result of 2.84~$\mu_{\rm B}$ \cite{Fukuhara1996} and than the expected value of 2.54~$\mu_{\rm B}$ for free Ce$^{3+}$ ion.
Although the CEF schemes are still controversial, the splitting energy between the excited and ground states were reported to be 200-300~K \cite{Kuwai2004, Ehm2007}, which is comparable to the Curie-Weiss fitting temperature range.
To determine effective moment precisely, the susceptibility measurements at higher temperature regions above 300~K are needed.
The Weiss temperature ${\it\Theta_{{a}(c)}}$ is evaluated as -206~K (-56~K), which is in agreement with the literature \cite{Fukuhara1996}.
For YbNi$_2$Ge$_2$, the effective moment is 4.5~(4.6)~$\mu_{\rm B}$, which is near 4.54~$\mu_{\rm B}$ of Yb$^{3+}$ ion, and ${\it\Theta_{{a}(c)}}$ is $-79$~K ($-156$~K).

CeNi$_2$Ge$_2$ has a maximum in $M/H$ at $T_{\chi}^{\rm max}\sim$~30~K for only $H~||~c$, whereas for YbNi$_2$Ge$_2$ a maximum at $T_{\chi}^{\rm max}\sim$~50~K appears for both $H$ directions.
In addition, the susceptibility of CeNi$_2$Ge$_2$ features an upturn at low temperatures.
Assuming the anomalous peak corresponds to a density of state of a quasiparticle at the Fermi energy, the increase in $\chi$ can be reproduced phenomenologically \cite{Aoki1997}.
The singularity in the density of states relates strongly to the non-Fermi-liquid behavior, which can also reproduce the temperature dependence observed for the specific heat.
In addition, the mode-mode coupling theory predicts critical exponents at the AFM quantum critical point giving a $\chi\propto -T^{1/4}$ and a $-T^{1/2}$ dependence for the specific heat. \cite{Hatatani1998}.
Although the evaluation of the exponent from Fig.~\ref{MT} is difficult because low-temperature data are absent, the upturn is a consequence of the proximity to the AFM critical point.
The upturn is strongly suppressed with fields, as will be discussed later.
In contrast, the susceptibility of YbNi$_2$Ge$_2$ monotonically decreases below $T_{\chi}^{\rm max}$, suggesting that YbNi$_2$Ge$_2$ is far from the magnetic instability, in agreement with a previous report \cite{Knebel2001}.

Note also that, by replacing the Ce site by Yb, the magnetic easy direction is switched from the $c$-axis to the basal plane although the magnetic anisotropy $\chi_a/\chi_c$ of less than 2 is quite small.
Such changes in magnetic anisotropy were also seen for CeRh$_2$Si$_2$ and YbRh$_2$Si$_2$ \cite{Settai1997, Custers2001}.
The anisotropy change between CeNi$_2$Ge$_2$ and YbNi$_2$Ge$_2$ may be due to their different CEFs \cite{Budko1999}.
With tetragonal symmetry, the magnetic anisotropy is mainly dominated by the $B_2^0O_2^0$ term of the CEF Hamiltonian, and the easy $c$-axis and the $ab$-plane are realized for the negative and positive $B_2^0$, respectively \cite{Gubbens2012}.
The CEF parameter $B_2^0$ is evaluated from the difference between ${\it\Theta_{a}}$ and ${\it\Theta_{c}}$ for $H||a(c)$, specifically, $B_2^0 = 10({\it\Theta_{a}-\Theta_{c}})/[3(2J-1)(2J+3)]$ \cite{Budko1999, Wang1971}.
Using the estimated ${\it\Theta_{{a}(c)}}$ from the Curie-Weiss fits, $B_2^0$ for CeNi$_2$Ge$_2$ and YbNi$_2$Ge$_2$ are respectively estimated as $-23$~K and 5~K, consistent with their magnetic anisotropy, i.e., the easy $c$-axis and $ab$-plane for the former and the latter, respectively.

Our main finding in this work is the discovery of the first example exhibiting metamagnetic-like nonlinear magnetization curves in both isomorphic Ce and Yb compounds having a PM ground state.
Figure \ref{MH} presents the magnetization curves $M(H)$ at 1.4~K for CeNi$_2$Ge$_2$ and YbNi$_2$Ge$_2$ for applied fields along $a$- and $c$-axes.
CeNi$_2$Ge$_2$ clearly exhibits metamagnetic behavior at $H_m$ = 45~T for $H~||~c$-axis, whereas for $H~||~a$-axis, $M$ monotonically increases up to the highest fields.
This anisotropic behavior, demonstrated here for the first time, is common among Ce PM metamagnets, CeRu$_2$Si$_2$ \cite{Haen1987} and CeFe$_2$Ge$_2$ \cite{Sugawara1999}.
Within experimental error, hysteresis is not observed over a field cycle.
Moreover, note the $M(H)$ behavior above $H_m$. 
In CeNi$_2$Ge$_2$, the linear extrapolation of the $M(H)$ curve above $H_m$ crosses the origin, which is also seen in powdered samples \cite{Fukuhara1996}.
This is in strong contrast to the isostructural Ce metamagnets, CeRu$_2$Si$_2$ \cite{Haen1987} and CeFe$_2$Ge$_2$ \cite{Sugawara1999}. 
The finite intercept in CeRu$_2$Si$_2$ may reflect the strength of the spin polarization.
When crossing $H_m$, CeNi$_2$Ge$_2$ seems to change its PM character to a weakly polarized spin state.

The hole analog YbNi$_2$Ge$_2$ exhibits magnetization upturn, which might be the first observation of metamagnetic behavior in a tetragonal Yb paramagnet.
Interestingly, the fields along both directions induce magnetization upturns, which may be consistent with the appearance of the peak in the susceptibility at almost identical $T_{\chi}^{\rm max}$, and therefore the same energy scale governs the maximum of the susceptibility and metamagnetism. 
The anomaly is clear to see for the easy magnetization $a$-axis at $H_m\sim40$~T than that for $H||c$. 
Contrary to CeNi$_2$Ge$_2$, the nonlinearity of magnetization is very weak as seen in the very broad peak of d$M$/d$H$, and $M$ does not tend to saturate at least up to 56~T.
The $J$~=7/2 degeneracy is suggested by the susceptibility and the Kadowaki-Woods ratio considering the degeneracy \cite{Yamaoka2010, Tsujii2005}.
Higher fields are necessary to saturate to the value 4~$\mu_{\rm B}$ for a free Yb$^{3+}$ ion.

\begin{figure}[t]
   \begin{center}
      \includegraphics[width=90mm]{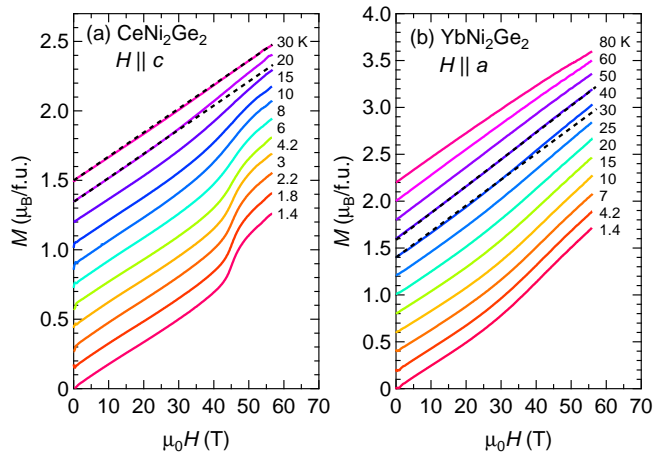}
   \end{center}
   \caption{(color online) Magnetization curves at various temperatures for (a) CeNi$_2$Ge$_2$ ($H||c$) and (b) YbNi$_2$Ge$_2$ ($H||a$).
   For clarity, the data are offset by (a) 0.15 and (b) 0.2~$\mu_{\rm B}$/f.u..
   The dashed lines are extrapolations of the linear field dependence of magnetization, suggesting the disappearance of metamagnetism near $T_{\chi}^{\rm max}$.
}
   \label{MH_Ts}
\end{figure}

\begin{figure}[t]
   \begin{center}
      \includegraphics[width=80mm]{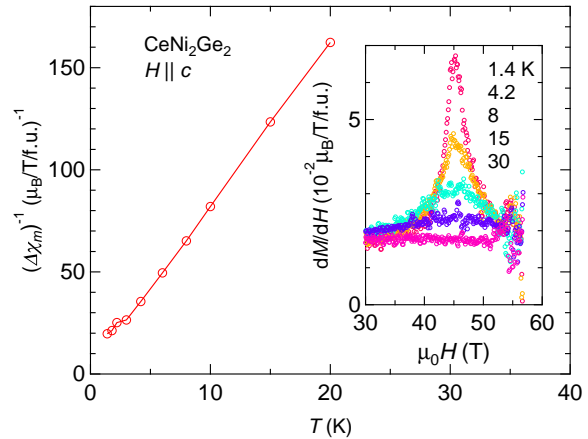}
   \end{center}
   \caption{(color online) Temperature dependence of the inverse peak height of the differential susceptibility of CeNi$_2$Ge$_2$ at $H_m$, $\Delta \chi_m$~=~$\chi_m-\chi_0$.
   The inset shows the temperature evolution of d$M$/d$H$.
   }
   \label{invChiT}
\end{figure}

Hereafter, we concentrate on the easy direction, specifically $H||c$ and $H||a$ for CeNi$_2$Ge$_2$ and YbNi$_2$Ge$_2$, respectively.
Fig.~\ref{MH_Ts} presents the $M(H)$ curves at various temperatures.
$H_m$ is insensitive to temperature, and with warming the $M$ anomaly becomes indistinct. 
The linearity of $M(H)$ (highlighted by the linear guide lines near $T_{\chi}^{\rm max}$) indicates the disappearance of metamagnetsm above $T_{\chi}^{\rm max}$.
In the inset of Fig.~\ref{invChiT}, the peak of $\chi$~=~d$M$/d$H$ in CeNi$_2$Ge$_2$ appears to disappear near $T_\chi^{\rm max}$.
The height of the differential susceptibility of CeNi$_2$Ge$_2$ at $H_m$ is determined by $\Delta\chi_m$~=~$\chi_m - \chi_0$, where $\chi_{m (0)}$ is the $\chi$ at $H$~=~$H_{m}$ (0.1~T). 
In contrast to the strong temperature dependence of $\chi_m$, $\chi_0$ exhibits very little dependence.
$\Delta\chi$ does not diverge at finite temperature, inferring a pseudo-metamganetic transition.
This behavior is also commonly observed in CeRu$_2$Si$_2$ and CeFe$_2$Ge$_2$ \cite{Haen1987, Sakakibara1995, Sugawara1999}.

\begin{figure}[t]
   \begin{center}
      \includegraphics[width=90mm]{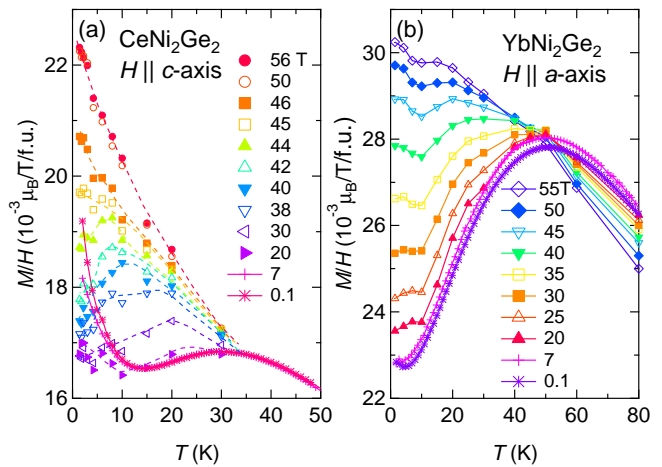}
   \end{center}
   \caption{(color online) Temperature dependence of $M/H$ of (a) CeNi$_2$Ge$_2$ ($H||c$) and (b) YbNi$_2$Ge$_2$ ($H||a$) at several constant fields.
      The broken lines in (a) are only a visual guide.
}
   \label{chi_T}
\end{figure}

To extract more details, we replotted $M/H$ as a function of temperature at various constant fields in Fig.~\ref{chi_T}, with the data from Fig.~\ref{MH_Ts}.
We first take a look at the characteristics for CeNi$_2$Ge$_2$.
With increasing field, the upturn at low temperatures is strongly suppressed and becomes constant at least above 20~T.
This indicates the recovery of the Fermi-liquid state.
$T_{\chi}^{\rm max}$ shifts to a lower temperature and tends towards 0~K as $H\rightarrow H_m$. 
A similar field evolution of the temperature dependence of $M/H$ was also reported for CeRu$_2$Si$_2$ \cite{Ishida1998}.
$M/H$ increases further with increasing field and saturates above $\sim$50~T.
Up to the highest studied fields $H/H_m\sim$~1.24, the suppression of $M/H$ at low temperature seen in CeRu$_2$Si$_2$ \cite{Ishida1998} is not observed.
Next, we take a look at YbNi$_2$Ge$_2$. 
The tendency is similar to CeNi$_2$Ge$_2$, i.e., the broad maximum shifts to lower temperature with increasing field.
In contrast to CeNi$_2$Ge$_2$, however, the broad maximum of $M/H$ in YbNi$_2$Ge$_2$ does not disappear even above $H_m = 40$~T.

For CeNi$_2$Ge$_2$, $T_{\chi}^{\rm max}$ is close in energy to the spectral linewidth of the AFM fluctuation obtained from inelastic neutron scattering \cite{Knopp1988, Fak2000, Kadowaki2003}; unfortunately, a similar measurement for YbNi$_2$Ge$_2$ is lacking.
Whether the magnetic and/or valence fluctuation exists in YbNi$_2$Ge$_2$ is important to know.
For CeNi$_2$Ge$_2$, as for CeRu$_2$Si$_2$, the suppression of the AFM fluctuation at $H_m$ drives $T_{\chi}^{\rm max}$ to zero \cite{Flouquet2004}.
Also low-energy spin fluctuations of around 0.6~meV were found to play an important role in the non-Fermi-liquid behavior \cite{Kadowaki2003}.
Therefore, this low-temperature behavior and the high-field pseudo-metamagnetic transition of CeNi$_2$Ge$_2$ are decoupled, as discussed in Ref.~\cite{Flouquet2005}.
A comparison of the thermal and magnetic Gr$\ddot{\rm u}$neisen parameters may resolve the above issues.

\begin{figure}[t]
   \begin{center}
      \includegraphics[width=90mm]{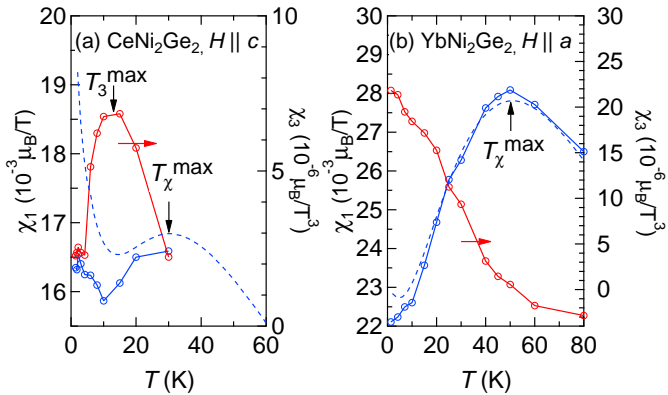}
   \end{center}
   \caption{(color online) Temperature dependence of $\chi_1$ (left axis) and $\chi_3$ (right axis) of (a) CeNi$_2$Ge$_2$ for $H||c$ and of (b) YbNi$_2$Ge$_2$ for $H||a$, respectively. 
Symbols and dotted lines are, respectively, fitted results and the $M/H$ measured at $\mu_{\rm 0}H=0.1$~T, respectively. 
}
   \label{chi3}
\end{figure}

The notable differences between CeNi$_2$Ge$_2$ and YbNi$_2$Ge$_2$ appear in the temperature dependence of the nonlinear susceptibility $\chi_3$.
The field expansion of the magnetization is written as $M(H) = \chi_1 H + \frac{1}{3!}\chi_3 H^3$, where $\chi_1$ is the uniform magnetic susceptibility and higher order terms are neglected.
Therefore, these values can be determined experimentally from the plot of $M/H$ vs $H^2$: the intercept and slope correspond to $\chi_1$ and $\chi_3$, respectively.
CeRu$_2$Si$_2$, for example, shows a maximum in both quantities: temperature $T_3^{\rm max}$ corresponds to a peak in $\chi_3$ that is below $T_{\chi}^{\rm max}$ \cite{Park1994}.

Figure~\ref{chi3} represents the temperature dependence of $\chi_1$ and $\chi_3$ for CeNi$_2$Ge$_2$ and YbNi$_2$Ge$_2$, respectively.
The consistency between $\chi_1$ obtained from the fit and $M/H$ data measured at $H$~=~0.1~T is rather good for YbNi$_2$Ge$_2$.
In contrast, the discrepancy is larger for CeNi$_2$Ge$_2$ because of the strong field-dependent non-Fermi-liquid behavior in $\chi(T)$ \cite{Aoki1997}.
For CeNi$_2$Ge$_2$, $\chi_3$ exhibits a maximum at $T_3^{\rm max}\sim$~13~K whereas for YbNi$_2$Ge$_2$ $\chi_3$ monotonically decreases with increasing temperature becoming negative at around $T_\chi^{\rm max}$.

Recently, Shivaram {\it et al}. pointed out that $T_{3}^{\rm max}$ is scaled by $T_{\chi}^{\rm max}$ in many heavy-fermion systems having a diverse type of metamagnetic transitions \cite{Shivaram2014}.
They proposed a simple two-level system model, i.e., an excited pseudospin of $S_z=\pm 1$ separated from the $S_z=0$ ground state by a gap $\Delta$ yielding the scaling $T_3^{\rm max}/T_{\chi}^{\rm max}\sim$~0.4.
The peak structures of $\chi_1$ and $\chi_3$ are dominated by a single energy scale $\Delta$, which is also related to $H_{m}$.
$T_3^{\rm max}$ of CeNi$_2$Ge$_2$ is near $T_{\chi}^{\rm max}/2$, following the scaling \cite{Shivaram2014}.

In striking contrast, YbNi$_2$Ge$_2$ does not show any peak structure in $\chi_3$.
The positive $\chi_3$ gradually decreases with increasing temperature becoming negative at around $T_{\chi}^{\rm max}$. 
The universality observed in many heavy-fermion compounds and CeNi$_2$Ge$_2$, i.e., $T_3^{\rm max}/T_{\chi}^{\rm max}\sim$~0.4 \cite{Shivaram2014}, is not valid for YbNi$_2$Ge$_2$.
The same characteristic behavior, however, were reported in nearly FM itinerant electron metamagnets YCo$_2$ \cite{Bloch1975} and TiBe$_2$ \cite{Mitamura1999}, which were not taken into consideration in the literature \cite{Shivaram2014}.
The sign change of $\chi_3$ at $T_{\chi}^{\rm max}$ was explained well using Landau theory including the spin fluctuations \cite{Yamada1993}. 
The Landau-type expansion uses $M$ as an order parameter, and shown to describe trends for the (nearly) FM systems well.
In contrast, for the AFM, the sublattice magnetization needs to be taken into account.
Quite recently, the metamagnetism of CeRu$_2$Si$_2$ and the related systems were phenomenologically explained using a Landau-type free energy for an AFM Ising systems with two sublattices \cite{Matsumoto2016}. 
The good description for YbNi$_2$Ge$_2$ using the usual Landau theory indicates that the metamagnetisms exhibited by CeNi$_2$Ge$_2$ and YbNi$_2$Ge$_2$ may be different in origin.
It was suggested that the FM interaction plays an important role in the magnetically ordered phase of YbNi$_2$Ge$_2$ under pressure \cite{Knebel2001}.
And thus, YbNi$_2$Ge$_2$ is located near FM critical point at ambient pressure, leading in the similarity to the nearly FM systems.
Moreover, most of the pressure-induced magnetic phases of Yb-based systems with ThCr$_2$Si$_2$ structure such as YbCu$_2$Si$_2$ \cite{Fernandez2011, Tateiwa2014}, YbIr$_2$Si$_2$\cite{Yuan2006} and YbRh$_2$Si$_2$ \cite{Plessel2003, Knebel2006} seem to be FM.
If the ordered phase above $p_c$ is FM, the field-induced first-order metamagnetic transition from PM to FM is expected near $p_c$ and in the PM phase \cite{Goto2001}.
Moving from $p_c$ to the PM side, the transition changes to crossover across the quantum critical endpoint, as found for UCoAl \cite{Aoki2011}.
For YbNi$_2$Ge$_2$, the PM phase is stabilized with decreasing pressure, and thus the metamagnetic crossover may take place at ambient pressure.
The broadness of the magnetization anomaly is because the FM critical point is located far away.
Pressure experiments can verify this scenario; pressure moves YbNi$_2$Ge$_2$ to $p_c$ and changes the metamagnetic anomaly from crossover to first order transition.
Of course, an experiment revealing magnetism above $p_c$ is most desired.
Because of a lack of other comparisons and experimental investigations of YbNi$_2$Ge$_2$, it is at present difficult to conclude whether metamagnetism has its origin in FM fluctuations.  

We discuss other alternative scenarios of the metamagnetic behavior in YbNi$_2$Ge$_2$.
The theory based on the Coqblin-Schrieffer model revealed a magnetization upturn for $J = 7/2$  and reproduced the metamgnetic behavior in YbCuAl with a single energy scale; it also explained the maximum in the temperature dependence of magnetic susceptibility and specific heat \cite{Hewson1983, Schlottmannn1989}.
In the calculation for $T$~=~0, the coefficient of $H^2$ term of $M/H$, $\chi_3$, was found to be positive in agreement with our results \cite{Hewson1983}.
Although there is no theoretical investigation of the temperature dependence of $\chi_3$, $\chi_3$~=~0 at $T_{\chi}^{\rm max}$ is at least expected with the disappearance of magnetization upturn. 
A further theoretical investigations of magnetization for $J$~=7/2 at finite temperatures is desired.
Metamagnetism in the valence crossover regime is theoretically known \cite{Watanabe2008}, in which divergences were seen not only for valence but also magnetic susceptibility at the field-induced valence quantum critical point and thereby initiating FM fluctuations.
Indeed, such metamagnetic behavior accompanied by a large valence change is confirmed experimentally in YbAgCu$_4$ \cite{YHMatsuda2012}.
This may give rise to a similarity between valence changed metamagnets and nearly FM itinerant metamagnets.    
The very broad $M(H)$ anomaly indicates weak magnetic and valence fluctuations; YbNi$_2$Ge$_2$ has a relatively high $p_c\sim$~5~GPa \cite{Knebel2001}.
The evaluation of the field dependences of valence and volume deserves further attention so as to understand the metamagnetic behavior observed in YbNi$_2$Ge$_2$.

Recently, field-induced Lifshitz transitions featuring magnetization anomalies have also been reported, for example in CeRu$_2$Si$_2$ \cite{Boukahil2014} and YbRh$_2$Si$_2$ \cite{Pfau2013, Pourret2013}.
The magnetization anomaly of the latter is a kink rather than a step \cite{Gegenwart2006}.
For both compounds, the effective mass is reduced across the transition.
Notably, the Lifshitz transition is not necessarily accompanied by a magnetization anomaly and a suppression of effective mass, as observed in CeIrIn$_5$ \cite{Aoki2016}.
For YbNi$_2$Ge$_2$ though, excluding the Lifshitz transition as the origin of the magnetization upturn is not possible at present.
In this regard, Fermi surface studies across $H_m$ gain some importance and urgency.

Also it is unclear at present whether the metamagnetic behavior of YbNi$_2$Ge$_2$ is a specific case or a more general of PM Yb systems having a tetragonal lattice.
Finding other examples such systems exhibiting similar properties and having a susceptibility maximum and easy-plane anisotropy would decide this issue.
YbCu$_2$Si$_2$, which is located near the FM phase separated by $p_c\sim$8~GPa \cite{Tateiwa2014}, has a susceptibility maximum at $T_{\chi}^{\rm max}\sim40$~K for $H~||~a$ \cite{Dung2009}.
At least up to 50~T, however, no clear metamagnetic behavior is observed in YbCu$_2$Si$_2$ in any direction, although the anisotropy $\chi_c/\chi_a\sim$~3 and thus CEF are different from that in YbNi$_2$Ge$_2$ \cite{Dung2009}.
Strong differences between CeNi$_2$Ge$_2$ and YbNi$_2$Ge$_2$ appear in the temperature dependence of $M/H$ near $H_m$ and in $\chi_3$ near $T_{\chi}^{\rm max}$.
To specify the characteristic features in YbNi$_2$Ge$_2$, determining whether other Yb metamagnets such as YbAgCu$_4$ \cite{Graf1995} and Yb$T_2$Zn$_{20}$ \cite{Hirose2011} display a maximum or sign change in $\chi_3$ would be of interest from a substitutional perspective.
Although the substitution effect of Ce for Yb are not yet clear, the CEF scheme affects the anisotropy and seems to determine magnetic and/or valence fluctuations. 
Theoretical investigation considering CEF is strongly desired.

\section{Conclusion}
From the magnetization measurements in a pulsed field, we have observed the first example of metamagnetic behavior in PM isomorphs CeNi$_2$Ge$_2$ and YbNi$_2$Ge$_2$.
The behavior in both is a crossover rather than a phase transition.
In contrast to a rather sharp pseudo-metamagnetic transition in CeNi$_2$Ge$_2$, the nonlinearity is very weak for YbNi$_2$Ge$_2$.
Similar to other PM systems, the pseudo-metamagnetic fields can be scaled by the temperature corresponding to the susceptibility maximum. 
Temperature dependence of the linear and nonlinear susceptibility shows strong contrasts between CeNi$_2$Ge$_2$ and YbNi$_2$Ge$_2$.
The differences seem to depend on whether the systems are located near a AFM or FM critical point.
Other possibilities, such as valence fluctuation and Lifshitz transition, are at present not excluded as the origin of the metamagnetic behavior of YbNi$_2$Ge$_2$.
These findings are sufficiently intriguing to stimulate further investigations of metamagnetism in these systems.

\begin{acknowledgments}
The authors are grateful to D. Aoki, H. Mitamura, Y. Shimizu, and S. Watanabe for fruitful discussions.
This work was partially supported by the MEXT of Japan Grants-in-Aid for Scientific Research (15K17700 and 16H04014).
\end{acknowledgments}

\end{document}